# All-Optical Adaptive Control of Quantum Cascade Random Lasers

S. Schönhuber[1,2], N. Bachelard[3], B. Limbacher[1,2] M.A. Kainz[1,2], A.M. Andrews[2,4], H. Detz[5], G. Strasser[2,4], J. Darmo[1,2], S. Rotter[3] and K. Unterrainer[1,2]

[1]*Photonics Institute, TU Wien, 1040 Vienna, Austria*
[2]*Center for Micro- and Nanostructures, TU Wien, 1040 Vienna, Austria*
[3]*Institute for Theoretical Physics, TU Wien, 1040 Vienna, Austria*
[4]*Institute for Solid-State Electronics, TU Wien, 1040 Vienna, Austria*
[5]*Central European Institute of Technology, Brno University of Technology, 61200 Brno, Czech Republic*

**Abstract**

Spectral fingerprints of molecules are mostly accessible in the terahertz (THz) and mid-infrared ranges, such that efficient molecular-detection technologies rely on broadband coherent light sources at such frequencies. THz Quantum Cascade Lasers can achieve octave-spanning bandwidths. However, their tunability and wavelength selectivity is often constrained by the geometry of their cavity. The recently introduced Quantum Cascade Random Lasers represent alternative sources of THz light, in which random scattering provides the required field confinement. The random resonator geometry greatly relaxes wavelength selectivity, thus producing radiation that is both spectrally broadband and collimated in the far-field. Yet, the intrinsic randomness of these devices' spectral emission strongly restricts the scope of their potential applications. In this work, we demonstrate the all-optical adaptive control and tuning of Quantum Cascade Random Lasers. The specificity of our random-laser sources is exploited to locally modify the system's permittivity with a near-infrared (NIR) laser beam and thereby substantially reconfigure the distribution of disorder. Using a spatial light modulator combined with an optimization procedure, the NIR illumination is spatially modulated to reshape the spectral emission and transform the initially multimode laser into a single mode source, which could be harnessed to perform self-referenced spectroscopic measurements. Moreover, we show that local NIR perturbations can be used to sense linear and nonlinear interactions amongst modes in the near field. Our work points the way towards the design of broadly tunable THz sources with greatly relaxed fabrication constraints.

# 1. Introduction

The sensing of well-defined molecular transitions is crucial in various domains such as in gas spectroscopy. Typically, these transitions correspond to narrow absorption lines, which, in order to be probed, rely on precisely designed THz Quantum Cascade Laser (QCL) sources that are manufactured through high-precision processes. Sweeping the frequency across the absorption spectrum of molecules requires moreover a tuning mechanism both to adjust the lasing wavelength and to correct built-in inaccuracies. Various technological solutions have been proposed to tune QCLs. Amongst them, slight frequency tuning was demonstrated using electrically-driven heat sinks to modify the refractive index of the active region[1]. External optical cavities coupled to a QCL device have also been employed to realize continuous tuning ranges from 12 to 50 GHz[2–5], while more substantial ranges of 67 and 240 GHz were achieved using micro-optomechanics[6,7]. Remarkably, an optical tuning mechanism relying on near-infrared (NIR) laser excitation was recently exploited by several groups to manipulate the output power and frequency of mid-infrared QCLs. A tuning range of several GHz was obtained for QCLs ridge resonators under such NIR illumination[8,9].

Recently, Quantum Cascade Random Lasers (QCRLs) that intrinsically feature multimode emission[10–14] were demonstrated. Random lasers are a class of lasers whose feedback mechanism relies on a disordered gain medium as opposed to conventional optical cavities[15–17]. The far field of these devices is collimated[11], which is particularly remarkable since typical QCLs are characterized by large output divergence originating from the diffraction caused by the mismatch between cavity dimensions' and wavelength. QCRLs are based on multiple scattering events at randomly placed air holes (etched through the active region) and the resulting lasing spectra do not contain equally-spaced modes. Therefore, QCRLs offer an ideal platform to perform wavelength tuning since their modes are not strongly predefined by the resonator geometry—the spectrum is not cavity inherited—and the lasing emission can be reshaped by modifying the scattering inside the active region. Moreover, gain competition and spatial hole burning are actively present in the selection or suppression of lasing modes in QCLs. As a result, QCRLs represent a promising playground to collect information about nonlinear interactions amongst modes.

In this article, we demonstrate the all-optical adaptive control and broad tuning of QCRLs. We conveniently make use of the fact that randomly placed air holes in QCRLs can couple an external NIR laser beam into the structure, which modifies the local permittivity and thereby changes the disorder distribution. Using a spatial light modulator combined with an iterative procedure, the spatial distribution of the NIR perturbation over the sample is optimized to robustly reshape the emission of the device, which, being initially multimode, is transformed into a tunable single mode lasing source. Finally, we explain that our approach points the way towards addressing major challenges in QCLs, such as the probing of near-field modal distributions or the investigation of nonlinear interactions amongst modes (e.g. spatial hole burning). This work opens the door for the conception of optically-controlled QCL sources, which are highly tunable while displaying reproducible and robust properties.

**2. Optical tuning of QCRLs through local NIR perturbation**

Our sample consists of an electrically-driven QCRL manufactured by etching holes into the active region (Fig. 1(a) and Methods). The holes are randomly distributed spatially and form a disordered permittivity, which generates THz emission that we collect through a Fourier-transform infrared spectrometer (FTIR) coupled *via* parabolic mirrors. A Ti:Sapphire femtosecond laser in the near-infrared (tuned at 813 nm) is focused on the chip and produces a perturbation spot (270 µm in diameter) that is spatially scanned across the surface of the device (1 mm in diameter) with a x-y translation stage.

In contrast to conventional QCLs, the specific structure of our QCRL enables the perturbation spot to locally tune the permittivity profile of the disorder in the laser cavity. The near-infrared (NIR) illumination of a semiconductor above its bandgap is responsible for the creation of electron-hole pairs, which translates into a change in carrier concentration and conductivity (Methods). Thus, the perturbation spot generates a photo-excited electron-hole plasma that modifies locally the permittivity of the semiconductor. With a 1 µm penetration depth of the field into the semiconductor, the NIR-induced refractive-index tuning usually reveals weak in conventional QCLs[13]. In our QCRL, however, the NIR spot illuminates a large area within the etched holes (Fig. 1(a)) and thereby generates substantial changes in the permittivity, which can be estimated through the variations of the device current. Under NIR illumination (i.e., in the presence of photoexcited carriers), the current through a constant bias voltage of 17.5 V (horizontal grey dashed line in Fig. 1 (b)) increases from 1.97 A to 2.13 A (vertical black and green dashed lines in Fig. 1(b), respectively). From the laser doping, the initial electron density is estimated to be close to $n_e = 3.8 \cdot 10^{15}$ cm$^{-3}$. At the exposed areas around the holes, the density increases up to $n_e = 2.4 \cdot 10^{16}$ cm$^{-3}$. This change in electron density shifts the plasma frequency of GaAs close to the emission frequency of the QCRL, thus small modifications lead to dramatic permittivity changes according to Drude's model (Methods). Initially at a value of $\epsilon = 12.9$, under NIR illumination the permittivity decreases close to $\epsilon = 8.0$.

While scanning the perturbation spot across the sample, the permittivity tuning materializes into strong modifications within the multimode QCRL spectrum. Fig. 1(c) displays the spectra of a non-perturbed ("w/o perturbation", black curve) and an example of perturbed QCRL ("NIR illumination", orange curve) illuminated at spot $(x, y) = (3,9)$, both of which are pumped under the same voltage (Methods). The QCRL is systematically scanned by the NIR beam across the whole chip and we observe significant modifications over the whole spectrum spanning over 350 GHz (12x12 sampling positions provided in Supplementary Figure S1). Although the NIR spot illuminates only about 8% of the surface, we observe in Fig. 1(c) the emergence ("1", dashed line arrow) and the attenuation ("2", dotted line arrow) of lasing modes at seemingly arbitrary frequencies, while simultaneously other modes vanish. This rearrangement of the lasing spectrum originates from modifications of the dielectric constant, which reshape the modes and their interactions as further explained in section 4. For specific locations of the NIR spot, the modes whose spatial distribution does not overlap with the perturbation remain unaffected ("3", solid line arrow).

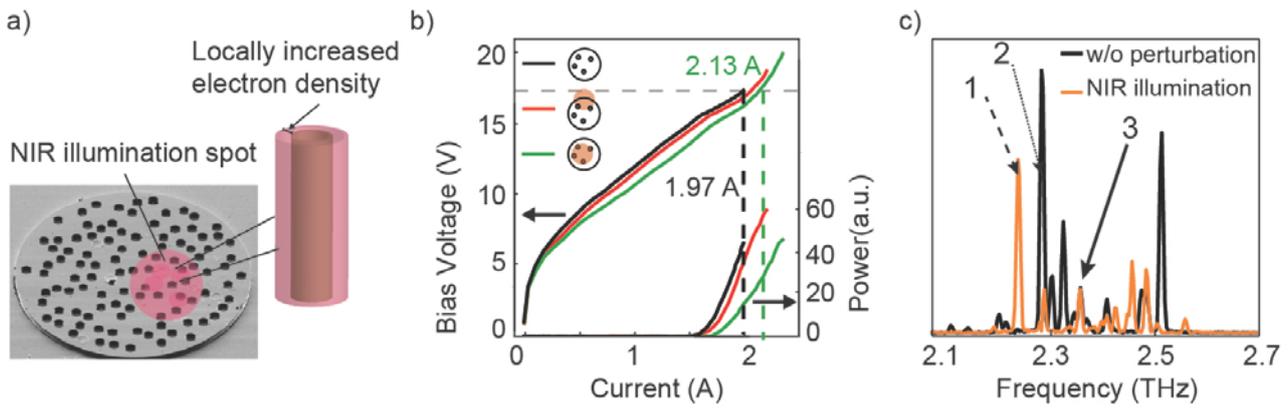

**Figure 1: Quantum Cascade Random Lasers under NIR illumination.** (a) Schematic representation of the tuning mechanism. SEM picture of a QCRL with an illustrated NIR laser spot (wavelength 813 nm, diameter 270 µm) that is scanned across the device. The QCRL (diameter 1 mm) is filled with holes (diameter 20 µm) for a filling fraction of 18 %. The NIR illumination is absorbed and creates electron-hole pairs within the holes, which leads to a local increase in electron density (red area around holes). (b) Light-Current-Voltage characteristics of the QCRL without perturbation (black), for a spot in the center (green) and at the edge (red) of the sample (see insets). Under a voltage bias of 17.5 V (horizontal grey dashed line), the current increases from 1.97 A (vertical black dashed line) to a maximum of 2.13 A (vertical green dashed line) due to photoexcited electrons. (c) Spectra measured in the absence of the perturbation ("w/o perturbation ", black curve) and in the presence of the NIR perturbation spot ("NIR illumination", orange curve) at position $(x, y) = (3,9)$. The mode structure is influenced by the NIR illumination. New modes emerge at seemingly arbitrary frequencies ("1", dashed line arrow), while existing modes are either attenuated ("2", dotted line arrow) or unaffected ("3", solid line arrow).

### 3. Mode selection

While changes in the position of the NIR laser spot already indicate the possibility of shaping the QCRL through optical means, more degrees of freedom in the control beam are required to realize the full potential of this approach and to achieve a complete control. Our strategy is to spread the NIR laser over the whole device area and implement an iterative procedure to spatially modulate the intensity of the beam. An adapted non-uniform distribution of the NIR intensity will then control the spectral emission of the QCRL. Inspired by former works on externally-controlled random lasing[18–20], we use a spatial light modulator to produce a spatially-optimized NIR perturbation pattern (Fig. 2(a) and Methods), which selectively transforms the initial emission of the device from multi-mode to single mode. Specifically, the QCRL is first electrically pumped until exhibiting three modes

(dashed lines in Figs. 2(b), (c) and (e)). We then use an optimization algorithm to reshape the NIR pattern through a nonlinear feedback loop, which aims to enhance the cost function $f = \frac{I_{opt}}{I_{ref}}$. Here, $I_{opt}$ stands for the intensity of the desired mode and $I_{ref}$ for the highest intensity amongst the remaining modes. Fig. 2(b) displays the spectrum recorded during the first step (Step 0) of the optimization procedure for the selection of a mode located at 2.3 THz (green circle), together with the corresponding spatial NIR distribution in inset (Methods). Fig. 2(c) shows the NIR pattern and the spectrum obtained after 54 iterations (Step 54), at which point we observe an almost single mode emission at frequency 2.3 THz with a cost function (identical to a rejection rate) close to $f = 20\ dB$. Fig. 2(d) displays the evolution of $1/f$ throughout the optimization and confirms the convergence of the procedure. The optimization can be applied to any mode within the emission range to produce a frequency-adapted single mode laser line. This point is illustrated in Fig. 2(e), in which the QCRL used in Figs. 2(b-d) is optimized to organize a single mode emission at a frequency of 2.32 THz (orange star).

Such a controlled THz source can find straightforward applications in domains like molecular spectroscopy, in which the lasing frequency has to match the absorption spectrum of respective elements. In particular, the QCRL emission range lies within the water-vapor absorption window (i.e. between 2.26 THz and 2.34 THz[21]) and, thus, this device can be used for the detection of molecules such as HCN or $NO_2$, which have dominant absorption lines at 2.319 THz and 2.298 THz, respectively[18]. Moreover, the QCRL can be controlled to lase at two or more different lasing frequencies simultaneously. Therefore, our platform is naturally suited to perform self-referenced spectroscopic measurements, which require two lasing modes (one frequency is used to characterize the absorption of the molecule while another acts as reference).

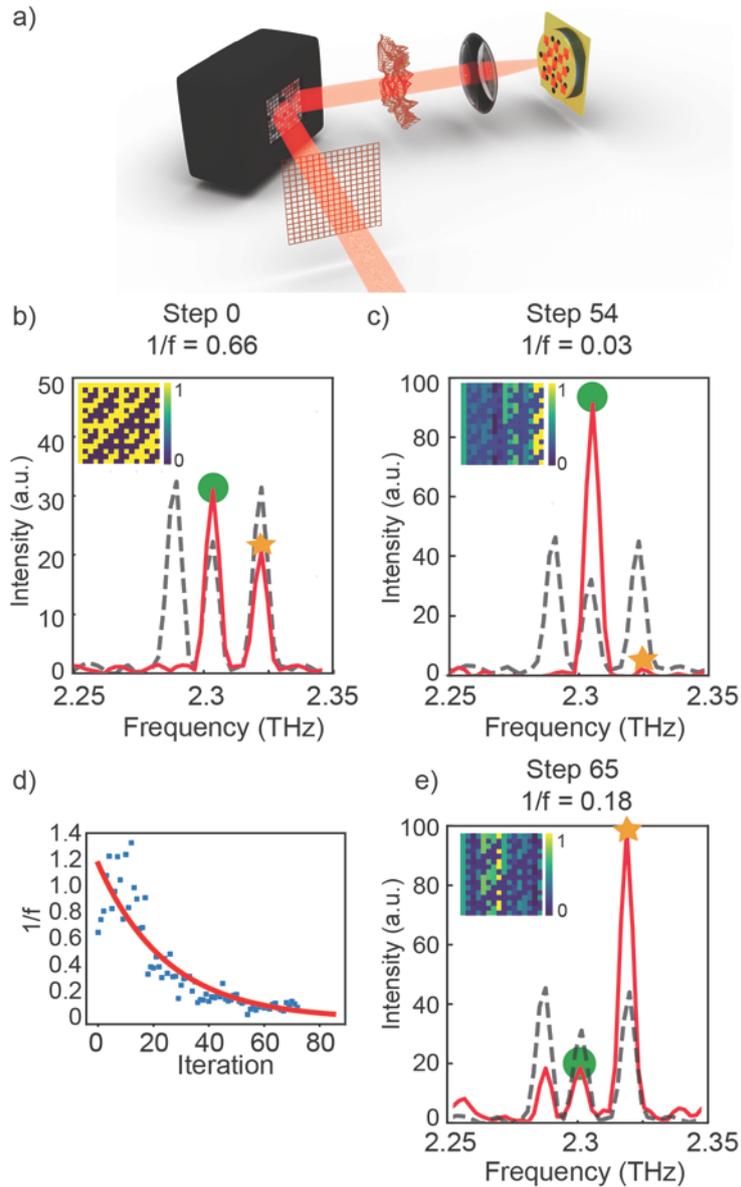

**Figure 2: Optically controlled mode selection in QCRLs.** (a) Setup performing a patterned NIR illumination on the QCRL. A spatial light modulator is used in reflection mode to spatially modulate the laser beam on the surface. The QCRL is coupled into the FTIR for spectral characterization (not shown). (b) Starting point for the iteration of the optimization routine applied to a mode at 2.3 THz (Step 0). An initial non-uniform pattern (Methods) is projected onto the QCRL (blue-yellow array in inset). The solid red curve corresponds to the spectrum collected under this non-uniform NIR illumination. The dashed gray curve shown here and in panels (c) and (e) was recorded when the NIR-perturbation is switched off. The green dot indicates the mode at 2.3 THz that the procedure intends to optimize, while the orange star indicates a second mode (at 2.32 THz) whose emission intensity we aim to reduce. (c) Final iteration of the optimization routine applied to the mode at 2.3 THz (Step 71). The algorithm converged towards a complex spatial modulation of the perturbation (blue-yellow array in inset) producing a single mode emission at 2.3 THz (green dots). (d) Evolution of the function $\frac{1}{f}$ for the optimization displayed in (b) and (c). (e) Final iteration of the optimization routine applied to a mode at 2.32 THz (step

65). The mode marked by an orange star (2.32 THz) is optimized by a non-uniform pattern (blue-yellow array in inset) obtained after 65 iterations.

## 4. Probing the mode structure

After illustrating the practical applicability of customized NIR illumination for laser control, we now demonstrate that this technique can also be envisioned for a spatial mapping of lasing modes. Since, in QCLs, the near field is typically not accessible and the far-field emission contains only partial information, accessing the structure of lasing modes remains a long-standing goal in the THz community. Here, we demonstrate that local perturbations can be exploited to sense both the spatial extent of the modes as well as their gain over the whole device. Specifically, we scan a localized NIR perturbation across the QCRL and observe dramatically different modulation behaviors amongst modes. These modes are tracked individually by selecting a particular frequency and storing its intensity in 12x12 matrices, in which each pixel corresponds to a spatial location of the NIR spot. Figs. 3(a) and (b) display the evolution while scanning the perturbation of a mode at 2.29 THz (label "2". in Fig. 1 (c)) and at 2.25 THz (label "1" in Fig. 1 (c)), respectively. The mode addressed in Fig. 3(a) emits for any position of the spot, but at different locations its intensity is strongly modulated while other modes are strongly impacted as well (see Supplementary Figure S2). In sharp contrast, the mode addressed in Fig. 3(b) does not emit in the spectrum collected in the absence of a perturbation. At very specific spot positions (e.g. yellow spot at $(x, y) = (5,1)$ or green spot at $(8,5)$) the mode starts lasing and the emissions of the remaining modes are strongly modified (see Supplementary Figure S3). These observations demonstrate that these two modes occupy rather different "spheres of action", which characterize the scanning area over which modes are strongly influenced by the NIR light and interact with each other.

To quantify whether the "spheres of action" of two modes are similar or complementary, we study mode mappings as displayed in Figs. 3(a) and (b) in terms of their cross-correlation. Fig. 3(c) presents the Pearson cross-correlation (Methods) between 14 modes, in which positive values indicate similar evolutions of the two selected modes under NIR illumination (simultaneous amplification or attenuation), while negative values characterize inverted behavior (simultaneous amplification of one mode and attenuation of the other). The inset panel displays the different correlations for the pairs of modes measured in Fig. 3(c) as a function of their frequency spacing, which emphasizes that modes close in frequency typically anticorrelate. Figs. 3(d) and (e) depict two modes—at 2.47 THz and 2.22 THz, respectively—associated with a high positive correlation value of 0.66 (red circle in Fig. 3(c)). Both modes display a similar response to NIR perturbations, which indicates that their modal intensity profiles do have a reinforcing overlap with each other. In contrast, Figs. 3(f) and (g) display two modes—at 2.47 THz and 2.50 THz, respectively—associated with a negative correlation value of -0.40 (green circle in Fig. 3(c)). Here, the overlap between the modal intensity profiles is destructive, which may be a signature of modal cross-saturation and hole-burning interactions in the competition for gain.

Due to the complex nature of modal interactions in random lasers, however, the correlations measured above can be the result of several mechanisms involved in the lasing emission. At first, modes are directly affected by changes in the refractive-index distribution induced by the NIR beam[19–23]. Through this modification certain modes initially below threshold can be brought above threshold and start lasing, while other modes, initially lasing, can be brought below threshold and be turned off. Alternatively, interactions amongst modes can be caused by nonlinear effects such as spectral and spatial hole burning that have been reported to play a major role in QCL mode selection[24–28]. These interactions are typically most pronounced for modes with a narrow frequency separation and a spatial overlap, under which conditions the modes enter a strong competition for the available gain. Here we observe indeed that pairs of modes with small frequency separations are in general associated with negative correlations (inset in Fig. 3(c)). The scarcity of positive correlations for pairs close in frequency suggests that modes with a spatial and spectral overlap interact strongly through nonlinear hole-burning. Amongst such modes typically the one with the lowest threshold depletes the gain accessible for all other modes, which thus cannot lase. Yet, drawing a definitive conclusion on the details of the mechanisms at play in this system will require further investigations (e.g. by applying our approach on simpler structures that are amenable to a numerical analysis).

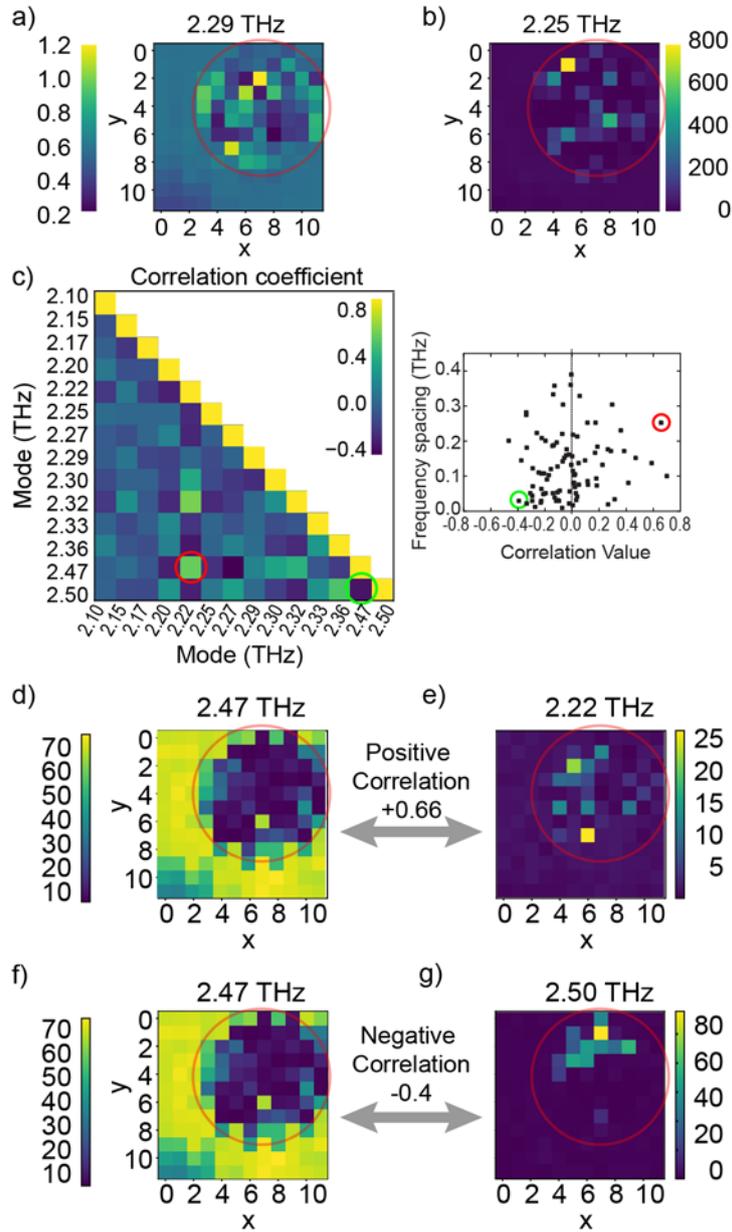

**Figure 3: Mode sensitivity map.** Spectral intensity (measured as a ratio with the unperturbed case) of individual modes at 2.29 THz (a) and 2.25 THz (b) while the perturbation spot is scanned across the QCRL. Each pixel in the displayed 12x12 matrices is associated to a spot location $(x, y)$ (the red circles mark the QCRLs' outer rim). (c) Pearson cross-correlation between the mapping of 14 different modes. The red circle indicates a strong positive correlation of 66% for the modes at 2.47 THz and 2.22 THz, while the green circle pinpoints a strong negative correlation of -40% for the modes at 2.47 THz and 2.50 THz. For each pair of modes, the scattered plot in inset reports the correlation between modes as a function of their frequency spacing. We observe a concentration of negative correlations for pairs close in frequency. The green and red circles correspond to the ones displayed in main panel. (d),(e) Mappings of the two modes at 2.47 THz and

2.22 THz, respectively, whose correlation is red-circled in (c). (f),(g) Mapping of the two modes at 2.47 THz and 2.50 THz, respectively, whose correlation is green-circled in (c).

## 5. Conclusion

We introduced a new tuning mechanism in QCRLs relying on an all-optical modulation of the active medium's permittivity. Through an algorithmic scheme, an adapted non-uniform distribution of NIR illumination across the device enables the formation of single mode and broadly tunable lasing. Our platform opens the way towards far-field collimated THz sources with an emission spectrum that is all-optically tunable *in-situ* and thus suitable for self-referenced spectroscopy and molecular detection. Moreover, our approach is intrinsically compatible with various QCL technologies and can thus be readily implemented to realize broadly configurable sources. The application of controlled local perturbations can also be envisioned for the investigation of a laser's complex nonlinear interactions such as spectral or spatial hole burning.

## METHODS

### Fabrication

Quantum cascade active regions are heterostructures formed by thin and alternating layers of different semiconductor materials. The created quantum wells form discrete energy levels that set the emission frequencies. Here, the active regions are realized in GaAs/Al$_{0.15}$Ga$_{0.85}$As, which are grown by molecular beam epitaxy and designed to provide an emission frequency around 2.3 THz (i.e. wavelength of 130 µm). The layer sequence starting from the injector barrier is 4.3/8.9/2.46/8.15/4.17/1/5/10 nm, the 5-nm well is homogeneously n-doped with a density of 6·10$^{16}$ cm$^{-3}$. The QCL module is repeated 340 times for a total thickness of 13 µm. In order to improve the electrical contact, a highly doped GaAs layer was grown at the bottom of the active region with a thickness of 100 nm. The active region and carrier substrate were covered with a 1 μm-thick gold layer, followed by an Au-Au thermo-compression bonding step. After lapping and wet etching, the gold-top contact layer was structured by a photo-lithography/lift-off process, acting as a self-aligned etch mask for the subsequent reactive ion etching process. The device is then mounted on a copper heat sink with indium and contacted using a wire-bonding technique.

### Light-Current-Voltage (L-I-V) and Spectrum measurements

The spectral measurements are performed using a Bruker Vertex 80 FTIR spectrometer with a resolution of 2.25 GHz. The emitted light is recorded with an attached pyroelectric deuterated triglycine sulfate (DTGS) detector. The sample is mounted in a liquid-helium-cooled-flow cryostat that is optically coupled to the spectrometer via parabolic mirrors. For L-I-V measurements, the light intensity $L$ is acquired by feeding the FTIR-detector output (internal DTGS detector) into a lock-in amplifier (Stanford Research Systems SR830) and measuring the signal at the modulation frequency 10 Hz. The current flowing through the device is measured with a coaxial current probe (Tektronix CT-1), which is connected to the output of the voltage pulser (HP8114A). The measurement data of the current $I$ and voltage $V$ are acquired using a digital oscilloscope (Tektronix DPO 3032).

**Near Infrared Illumination and Light Shaping.**

NIR illumination is provided by a Ti:Sapphire Mira 9000 laser system with both tunable wavelength and tunable output power, which operates in continuous mode. For an efficient perturbation of the QCRL, the wavelength is set to 813 nm with an intensity of 250 mW. The wavefront of the laser beam is spatially modulated in the Fourier space with a phase mask (generated by a spatial light modulator in the reflection mode) and then projected onto the QCL through a lens that creates the corresponding real space pattern. The phase mask is calculated using the Gerchberg-Saxton algorithm[28]

The NIR illumination of a semiconductor heterostructure leads to the excitation of electron-hole pairs. The NIR light is coupled into the structure through holes in the top gold layer. Here, we assume a simple rectangular excitation profile with a thickness of 1 µm (corresponding to the absorption length), which allows the estimation of the local electron density. The change in permittivity is calculated by applying Drude's mode[29]

$$\epsilon_{THz} = \epsilon \left(1 - \frac{\omega_P^2}{\omega^2}\right)$$

Here, we consider a permittivity of GaAs $\varepsilon = 12.9$, an angular frequency $\omega = 2\pi \cdot 2.3 \cdot 10^{12}\ Hz$ and the plasma frequency

$$\omega_P^2 = \frac{ne^2}{\epsilon_0 \epsilon m^*}$$

which contains the respective electron density $n$, the electron charge $e$, the vacuum permittivity $\varepsilon_0$ and the effective electron mass $m^* = 0.067 \cdot 9.11 \cdot 10^{-31}$ kg.

**Optimization Algorithm**

The pattern for mode selection is optimized using a Broyden-Fletcher-Goldfarb-Shanno gradient-based algorithm to find an optimum solution[30,31].

The NIR patterns are projected along an unmodified binary Hadamard basis similar to the one used in former works[18]. The cost-function of the algorithm is the divisor of $I_{ref}/I_{min}$ (described in the main text), which is iteratively fed back into the algorithm.

**Mode Interaction**

The interaction between individual mode-sensitivity maps X and Y is quantified by Pearson's correlation coefficient, which is defined as

$$\rho := \frac{Cov(X,Y)}{\sigma_x \sigma_y}$$

in which Cov(X,Y) stands for the covariance of X and Y and $\sigma_x$ and $\sigma_y$ to the associated variances.

## Acknowledgements

We thank G. Reider for very fruitful discussions. Financial support received by the Vienna Science and Technology Fund (WWTF) through Project No. MA09-030, by the Austrian Science Fund (FWF) through Projects No. F25 (SFB IR-ON), No. F49 (SFB NextLite), No. W1210 (DK CoQuS), and No. W1243 (DK Solids4Fun). H.D. is an APART Fellow of the Austrian Academy of Sciences. This project has received funding support from the European Union's Horizon 2020 research and innovation program under the Marie Skłodowska-Curie grant agreement No.840745. S.S. and M.K. performed the experiments, S.S, B.L. and N.B. implemented the optimization algorithm. H.D. A.M. A. and G.S. grew the quantum cascade heterostructure. S.R. J.D., K.U. and N.B. initiated and supervised the project at all stages. S.S. and N.B. wrote the first manuscript draft. All authors discussed the results and contributed to final manuscript version.